# Transmission line parameter identification using PMU measurements

Di Shi[1,*,†], Daniel J. Tylavsky[1], Kristian M. Koellner[2], Naim Logic[2], David E. Wheeler[2]

[1]*School of Electrical Computer and Energy Engineering, Arizona State University, Tempe, AZ 85287-5706, USA*

[2]*Salt River Project, Phoenix, AZ 85072-2025, USA*

SUMMARY

Accurate knowledge of transmission line (TL) impedance parameters helps to improve accuracy in relay settings and power flow modeling. To improve TL parameter estimates, various algorithms have been proposed in the past to identify TL parameters based on measurements from Phasor Measurement Units (PMU's). These methods are based on the positive sequence TL models and can generate accurate positive sequence impedance parameters for a fully-transposed TL when measurement noise is absent; however these methods may generate erroneous parameters when the TL's are not fully-transposed or when measurement noise is present. PMU field-measure data are often corrupted with noise and this noise is problematic for all parameter identification algorithms, particularly so when applied to short transmission lines. This paper analyzes the limitations of the positive sequence TL model when used for parameter estimation of TL's that are untransposed and proposes a novel method using linear estimation theory to identify TL parameters more reliably. This method can be used for the most general case: short/long lines that are fully transposed or untransposed and have balanced/unbalance loads. Besides the positive/negative sequence impedance parameters, the proposed method can also be used to estimate the zero sequence parameters and the mutual impedances between different sequences. This paper also examines the influence of noise in the PMU data on the calculation of TL parameters. Several case studies are conducted based on simulated data from ATP to validate the effectiveness of the new method. Through comparison of the results generated by this novel method and several other methods, the effectiveness of the proposed approach is demonstrated.

## 1. INTRODUCTION

Accurate transmission line (TL) impedance parameters are of great importance in power system operations for all types of system simulations, such as transient stability, state estimation etc., and are used as the basis for protective relay settings. TL parameters in the past have been estimated by engineers based on the tower geometries, conductor dimensions, estimates of actual line length, conductor sag, and other factors [1]. These calculated parameters are based on assumptions and approximations.

With the development of the PMU technology, synchronized phasors offer the possibility of allowing accurate estimation of transmission line parameters. Accurate knowledge of TL impedance parameters helps to:

- Improve accuracy in relay settings.
- Improve post-event fault location and thus lead to a quicker restoration of the systems.
- Improve transmission-line modeling for system simulations, such as state estimation calculations.
- Determine when the model for a transmission line in the centralized database has not kept pace with modifications to that transmission line, such as the insertion of series capacitors, extension of the line,

*Correspondence to: Di Shi, Department of Electrical Engineering, Arizona State University, Tempe, AZ 85287-5706, USA
†E-mail: Di.Shi@asu.edu

re-conductoring of the line, etc.

Several methods have been proposed in the past to identify TL parameters using PMU measurements [3-6]. One two-port ABCD parameter based method is proposed in [3]. This method utilizes two samples of synchronized measurements[2] from each terminal of the TL to identify the ABCD parameters; from these chain parameters the impedance parameters can be calculated. In this work, we refer to this method as the "two measurement method."

Another simpler method proposed in [4] requires only one sample from the two terminals of a TL to calculate the TL impedances directly; this method is henceforth referred to as the "single measurement method" by the authors. Both methods in [3] and [4] have drawbacks. First they do not perform well when there is noise in the phasor measurements. Second, these methods are based on the positive sequence TL model which is suitable only for fully transposed TL's; when the TL's are untransposed or not fully transposed, applying these methods will lead to considerable errors in the calculated parameters.

Reference [5] proposes a method based on the distributed TL model and uses nonlinear estimation theory to generate an optimal estimator of the fault location and TL parameters. With redundant sampling of measurements, this method reduces the effects of random noise to errors in the calculated parameters, but this method still has limitation since it is based on the positive sequence TL model only and neglects the unbalance in the system.

It is well known that for fully transposed TL's, the three sequence networks are completely decoupled and the positive sequence impedance parameters are determined by only the positive sequence voltages and currents. However, for untransposed TL's or TL's that are not fully transposed, the three sequence networks will be mutually coupled; using only the positive sequence measurements in these cases to estimate the positive sequence parameters will generate inaccurate parameter estimates.

In this paper, the limitation of positive sequence TL model for the purposes of parameter estimation is addressed and a novel method is proposed. The new method can be used to identify TL parameters for the most general case: short/long transposed/untransposed lines with balanced/unbalanced load conditions. The method can be used to calculate the parameters for both short TL's with the nominal pi model and long TL's with an equivalent pi model. The new method is based on the linear estimation theory and employs multiple PMU measurements; it generates satisfactory results even when the measurements are corrupted with noise.

This paper is organized as follows: Section 2 introduces the single and double measurement methods. Derivation of the new method is presented in section 3 along with a discussion of the limitations of the using the positive sequence TL model for parameter estimation. In section 4, several case studies based on simulated data from ATP are introduced and the effectiveness of the new method is validated. The main conclusions of the work are summarized in section 5.

## 2. THE SINGLE AND DOUBLE MEASUREMENT METHODS

A general three phase TL model is shown below in Figure 1, where

$$\overline{I}_{abc}^{S(R)} = [I_a^{S(R)} I_b^{S(R)} I_c^{S(R)}]^T, \overline{U}_{abc}^{S(R)} = [U_a^{S(R)} U_b^{S(R)} U_c^{S(R)}]^T,$$

$$Y_{abc} = \begin{bmatrix} Y_{aa} & Y_{ab} & Y_{ac} \\ Y_{ab} & Y_{bb} & Y_{bc} \\ Y_{ac} & Y_{bc} & Y_{cc} \end{bmatrix}, Z_{abc} = \begin{bmatrix} Z_{aa} & Z_{ab} & Z_{ac} \\ Z_{ab} & Z_{bb} & Z_{bc} \\ Z_{ac} & Z_{bc} & Z_{cc} \end{bmatrix}.$$

[2] By one sample of measurements we mean the time-domain synchronized phasor measurements (containing the same time stamp) of voltage and current taken from all three phases at both ends of a transmission line.

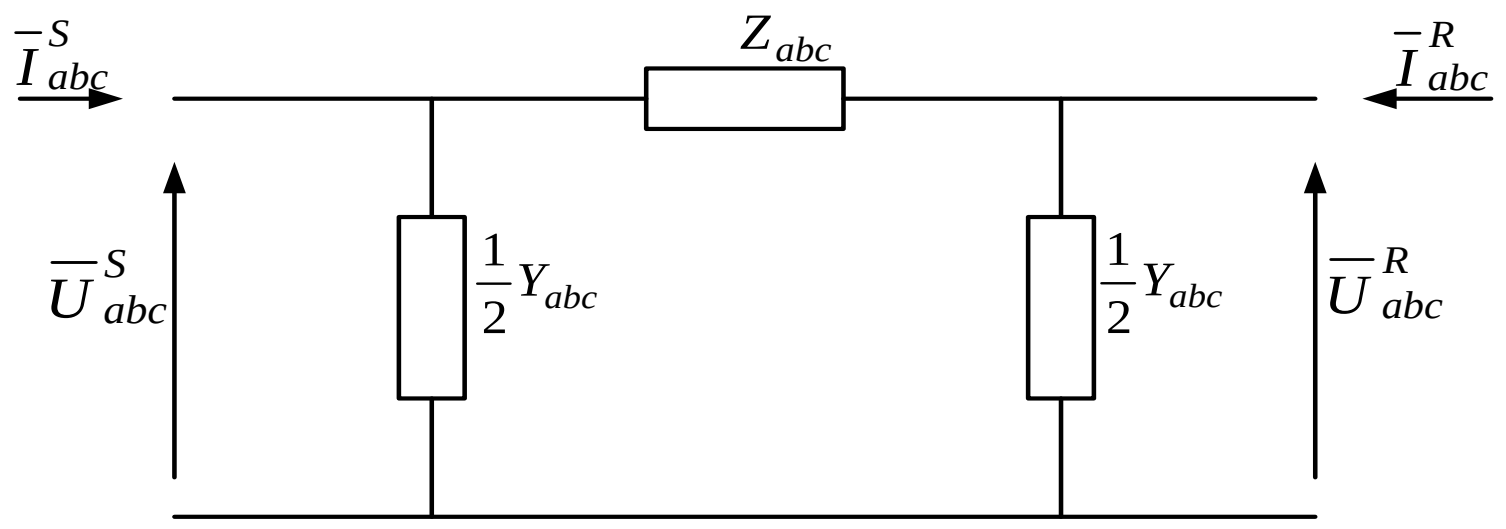

Figure 1. 3-phase transmission line model

From nodal analysis, we can write the following two equations (1) and (2) for Figure 1:

$$\overline{U}^{S}_{abc} - \overline{U}^{R}_{abc} = Z_{abc} \cdot (\overline{I}^{S}_{abc} - \frac{Y_{abc}}{2} \cdot \overline{U}^{S}_{abc}) \quad (1)$$

$$\overline{I}^{S}_{abc} + \overline{I}^{R}_{abc} = \frac{Y_{abc}}{2} \cdot (\overline{U}^{S}_{abc} + \overline{U}^{R}_{abc}) \quad (2)$$

The voltage and current variables in the equations above are all phase-frame-of-reference quantities. In order to transform these quantities to sequence-frame-of-reference quantities, we need to apply the phase-to-sequence transformation matrix, which is defined as:

$$A = \begin{bmatrix} 1 & 1 & 1 \\ 1 & a^2 & a \\ 1 & a & a^2 \end{bmatrix} \text{ where } a = e^{j120^0}$$

The following relationships then hold between the phase quantities and sequence quantities:

$$\overline{U}^{S(R)}_{abc} = A \cdot \overline{U}^{S(R)}_{012} \qquad \overline{I}^{S(R)}_{abc} = A \cdot \overline{I}^{S(R)}_{012}$$

$$Z_{012} = A^{-1} Z_{abc} A \qquad Y_{012} = A^{-1} Y_{abc} A$$

By multiplying both sides of (1) and (2) by the matrix $A^{-1}$, these equations can be rewritten as:

$$\overline{U}^{S}_{012} - \overline{U}^{R}_{012} = (A^{-1} Z_{abc} A) \cdot (\overline{I}^{S}_{012} - \frac{1}{2} \cdot A^{-1} Y_{abc} A \cdot \overline{U}^{S}_{012}) = Z_{012} \cdot (\overline{I}^{S}_{012} - \frac{Y_{012}}{2} \cdot \overline{U}^{S}_{012}) \quad (3)$$

$$\overline{I}^{S}_{012} + \overline{I}^{R}_{012} = \frac{1}{2} \cdot (A^{-1} Y_{abc} A) \cdot (\overline{U}^{S}_{012} + \overline{U}^{R}_{012}) = \frac{1}{2} \cdot Y_{012} \cdot (\overline{U}^{S}_{012} + \overline{U}^{R}_{012}) \ . \quad (4)$$

For fully transposed transmission lines, there are two independent components in the phase series impedance matrix, $Z_{mutual}$ and $Z_{self}$, since $Z_{aa} = Z_{bb} = Z_{cc} = Z_{self}$ and $Z_{ab} = Z_{bc} = Z_{ac} = Z_{mutual}$. The phase-frame-of-reference series impedance matrix has the following form:

$$Z_{abc} = \begin{bmatrix} Z_{self} & Z_{mutual} & Z_{mutual} \\ Z_{mutual} & Z_{self} & Z_{mutual} \\ Z_{mutual} & Z_{mutual} & Z_{self} \end{bmatrix} \quad (5)$$

As a result, the sequence series impedance matrix for a fully transposed line is well known to be diagonal of the form:

$$Z_{012} = A^{-1} Z_{abc} A = \begin{bmatrix} Z_0 & 0 & 0 \\ 0 & Z_1 & 0 \\ 0 & 0 & Z_2 \end{bmatrix} \quad (6)$$

where $Z_{ij}$ $(i = j)$ is the self impedance for each sequence network

The sequence shunt susceptance matrix is also found to be diagonal following the same derivation. Therefore, each of (3) and (4) can be broken up into three independent equations, of which the positive sequence equations are:

$$. \quad U_1^S - U_1^R = Z_{11} \cdot (I_1^S - \frac{Y_{11}}{2} \cdot U_1^S) \quad (7)$$

$$I_1^S + I_1^R = \frac{1}{2} \cdot Y_{11} \cdot (U_1^S + U_1^R) \quad (8)$$

This means that the sequence networks are fully decoupled and the positive sequence impedance parameters are only determined by the positive sequence voltages and currents. Hence, as is well known, for fully transposed TL's and perfect measurements (without noise), we can get accurate positive sequence impedance parameters using only the positive sequence phasor measurements even if the currents flowing through the transmission line are unbalanced.

As mentioned earlier, two parameter estimation methods based on this positive sequence model have been proposed; the single measurement and double measurement methods. The single measurement method [4] is derived by solving (7) and (8) for $Z_{11}$ and $Y_{11}$ . The impedance parameters can be obtained following the two equations:

$$Z_{11} = \frac{(U_1^S)^2 - (U_1^R)^2}{I_1^S U_1^R - I_1^R U_1^S} \quad (9)$$

$$Y_{11} = \frac{2(I_1^S + I_1^R)}{U_S^1 + U_R^1} \quad (10)$$

Compared with the single measurement method, the method proposed in [3] (referred to in this paper as the double measurement method) utilizes two sample of measurements and calculates the impedance parameters using a two step procedure. First the ABCD parameters of the TL are estimated using the following chain parameter equations, which is based on the two-port network model [1]:

$$U_{11}^S = AU_{11}^R - BI_{11}^R \quad (11)$$

$$I_{11}^S = CU_{11}^R - DI_{11}^R \quad (12)$$

$$U_{12}^S = AU_{12}^R - BI_{12}^R \quad (13)$$

$$I_{12}^S = CU_{12}^R - DI_{12}^R \quad (14)$$

where

$U_{11}^S, U_{11}^R, I_{11}^S, I_{11}^R$ Positive sequence phasor measurements from sample #1

$U_{12}^S, U_{12}^R, I_{12}^S, I_{12}^R$ Positive sequence phasor measurements from sample #2

Solving the four complex equations (11)~(14) with four unknowns based on Cramer's Rule gives:

$$A = \frac{I_{11}^{R} U_{12}^{S} + I_{12}^{R} U_{11}^{S}}{\det} \tag{15}$$

$$B = \frac{U_{11}^{S} U_{12}^{S} - U_{11}^{R} U_{12}^{S}}{\det} \tag{16}$$

$$C = \frac{I_{11}^{S} I_{12}^{R} - I_{12}^{S} I_{12}^{R}}{\det} \tag{17}$$

$$D = \frac{I_{11}^{S} V_{12}^{R} - I_{12}^{S} V_{11}^{R}}{\det} \tag{18}$$

where

$$\det = V_{11}^{R} I_{12}^{R} - V_{12}^{R} I_{11}^{R}$$

Once the chain parameters are calculated, the impedance parameters can be calculated directly from the following relationships:

$$A = 1 + 0.5 \cdot Y_{11} \cdot Z_{11} \tag{19}$$

$$B = Z_{11} \tag{20}$$

$$C = Y_{11} \cdot (1 + 0.25 \cdot Y_{11} \cdot Z_{11}) \tag{21}$$

$$D = 1 + 0.5 \cdot Y_{11} \cdot Z_{11} \tag{22}$$

## 3. AN OPTIMAL PARAMETER ESTIMATION METHOD FOR UNTRANSPOSED LINES

The limitation with the single and double measurement methods (as we will show later) is that they are sensitive to noise [4], and are found lacking when the transmission line is untransposed and operated under unbalanced conditions. One objective of this paper is to arrive at an optimal parameter identification method that not only is less sensitive to noise, but also is applicable to the case where the TL's are untransposed.

Following a derivation similar to that of the previous section, when the transmission line is not fully transposed, the sequence impedance matrix has the following form:

$$Z_{012} = A^{-1} Z_{abc} A = \begin{bmatrix} Z_0 & Z_{01} & Z_{02} \\ Z_{10} & Z_1 & Z_{12} \\ Z_{20} & Z_{21} & Z_2 \end{bmatrix} \tag{23}$$

where $Z_{ij}(i \neq j)$ is the mutual impedance between different sequence networks.

This result is well known: For a TL that is untransposed or not fully transposed, the sequence impedance matrix is not diagonal and there is mutual coupling between the three sequence networks. As a result, we will have to take into account of the effects of negative and zero sequence components when we calculate the positive sequence impedances under balanced/unbalanced loading conditions.

### *3.1 Description of the proposed model*

In the 3-phase TL nominal/equivalent pi model, the shunt admittance matrix is comprised of two parts: the shunt conductance (real part) and the shunt susceptance (imaginary part). Compared to shunt susceptance, shunt conductance is negligible, and thus, in the proposed method, it is neglected. Equation (1) and (2) can be written as:

$$\overline{U}_{abc}^{S} - \overline{U}_{abc}^{R} = Z_{abc} \cdot (\overline{I}_{abc}^{S} - j \cdot \frac{1}{2} \cdot B_{abc} \cdot \overline{U}_{abc}^{S}) \tag{24}$$

$$\overline{I}_{abc}^{S} + \overline{I}_{abc}^{R} = j \cdot \frac{1}{2} \cdot B_{abc} \cdot (\overline{U}_{abc}^{S} + \overline{U}_{abc}^{R}) \tag{25}$$

For a transmission line, the impedance matrix, $Z_{abc}$, is always symmetrical and can be written as:

$$Z_{abc} = \begin{bmatrix} Z_a & Z_{ab} & Z_{ac} \\ Z_{ab} & Z_b & Z_{bc} \\ Z_{ac} & Z_{bc} & Z_c \end{bmatrix} \tag{26}$$

Since the inverse matrix of $Z_{abc}$ is symmetrical, denote $Z_{abc}^{-1}$ as $y_P$ to obtain:

$$y_P = \begin{bmatrix} y_a & y_{ab} & y_{ac} \\ y_{ab} & y_b & y_{bc} \\ y_{ac} & y_{bc} & y_c \end{bmatrix} \tag{27}$$

For equation (24), multiplying both sides by $Z_{abc}^{-1}$ generates:

$$y_P \cdot (\overline{U}_{abc}^{S} - \overline{U}_{abc}^{R}) = \overline{I}_{abc}^{S} - j \cdot \frac{1}{2} \cdot B_{abc} \cdot \overline{U}_{abc}^{S} \tag{28}$$

Rewriting equation (28) and (25) into matrix format:

$$\begin{bmatrix} y_a & y_{ab} & y_{ac} \\ y_{ab} & y_b & y_{bc} \\ y_{ac} & y_{bc} & y_c \end{bmatrix} \begin{bmatrix} \Delta U_a \\ \Delta U_b \\ \Delta U_c \end{bmatrix} = \begin{bmatrix} I_a^S \\ I_b^S \\ I_c^S \end{bmatrix} - j \frac{1}{2} \cdot \begin{bmatrix} B_a & B_{ab} & B_{ac} \\ B_{ab} & B_b & B_{bc} \\ B_{ac} & B_{bc} & B_c \end{bmatrix} \begin{bmatrix} U_a^S \\ U_b^S \\ U_c^S \end{bmatrix} \tag{29}$$

$$\begin{bmatrix} \Delta I_a \\ \Delta I_b \\ \Delta I_c \end{bmatrix} = j \frac{1}{2} \cdot \begin{bmatrix} B_a & B_{ab} & B_{ac} \\ B_{ab} & B_b & B_{bc} \\ B_{ac} & B_{bc} & B_c \end{bmatrix} \begin{bmatrix} U_a^S + U_a^R \\ U_b^S + U_b^R \\ U_c^S + U_c^R \end{bmatrix} \tag{30}$$

where

$$\Delta U_x = U_a^S - U_a^R \; (x = a, b, or \quad c)$$

$$\Delta I_x = I_x^S + I_x^R \; (x = a, b, or \quad c)$$

Further expanding equation (29) and (30) yields 6 complex equations:

$$y_a \Delta U_a + y_{ab} \Delta U_b + y_{ac} \Delta U_c = I_a^S - j \frac{1}{2} \cdot (B_a U_a^S + B_{ab} U_b^S + B_{ac} U_c^S) \tag{31}$$

$$y_{ab} \Delta U_a + y_b \Delta U_b + y_{bc} \Delta U_c = I_b^S - j \frac{1}{2} \cdot (B_{ab} U_a^S + B_b U_b^S + B_{bc} U_c^S) \tag{32}$$

$$y_{ac} \Delta U_a + y_{bc} \Delta U_b + y_c \Delta U_c = I_c^S - j \frac{1}{2} \cdot (B_{ac} U_a^S + B_{bc} U_b^S + B_c U_c^S) \tag{33}$$

$$\Delta I_a = j \frac{1}{2} \cdot \left[ B_a (U_a^S + U_a^R) + B_{ab} (U_b^S + U_b^R) + B_{ac} (U_c^S + U_c^R) \right] \tag{34}$$

$$\Delta I_b = j \frac{1}{2} \cdot \left[ B_{ab} (U_a^S + U_a^R) + B_b (U_b^S + U_b^R) + B_{bc} (U_c^S + U_c^R) \right] \tag{35}$$

$$\Delta I_c = j \frac{1}{2} \cdot \left[ B_{ac} (U_a^S + U_a^R) + B_{bc} (U_b^S + U_b^R) + B_c (U_c^S + U_c^R) \right] \tag{36}$$

In equations (31)~(36), noticing $y_x (x = a, b, c, ab, bc, or \quad ac)$ is a complex number, we define: $y_x = G_x + j \cdot T_x (x = a, b, c, ab, bc, or \quad ac)$ For the purpose of obtaining an optimal estimate of the *y* and *B* parameters in (31)~(36), we expand these 6 complex equations into 12 real equations. Due to limited space, these 12 real equations are not listed here but can be found in APPENDIX A and the detailed derivations are presented in [7]. The discussion below is based on these 12 real equations.

In order to generate a simple and uniform expression for the problem, the following definitions are made:

Define *X* to be the measurement vector, which is known and can be calculated from the PMU measurements. That is,

$$\begin{aligned} X &= [x_1, x_2, ..., x_{24}]^T \\ &= [\text{Re}(U_a^S), \text{Im}(U_a^S), \text{Re}(U_b^S), ..., \text{Re}(U_a^R), \text{Im}(U_a^R), ..., \text{Re}(I_a^S), \text{Im}(I_a^S), ..., \text{Re}(I_a^R), \text{Im}(I_a^R) ...]^T \end{aligned} \quad (37)$$

where *Re(.)* and *Im(.)* yield the real and imaginary part of the input argument, respectively.

Define $\theta$ to be the unknown parameter vector, which is composed of the unknown impedance parameters. That is,

$$\theta = [\theta_1, \theta_2, ..., \theta_{24}]^T = [G_a, T_a, G_b, T_b, ..., G_{ab}, T_{ab}, G_{bc}, ..., T_{ac}, B_a, B_b, B_c, B_{ab}, B_{bc}, B_{ac}]^T \quad (38)$$

From the definition (37), we may further define a measurement vector:

$$Z = [x_{19}, x_{21}, x_{23}, x_{19} - x_{13}, x_{21} - x_{15}, x_{23} - x_{17}, x_{20}, x_{22}, x_{24}, x_{14} - x_{20}, x_{16} - x_{22}, x_{18} - x_{24}]^T \quad (39)$$

Based on the definitions above, we further arrange these 12 equations into matrix format as:

$$Z = H \cdot \theta^T \quad (40)$$

where $H$ is a matrix formulated from the 12 equations referenced above and contains measurements. And the measurement vector, Z, contains PMU voltage and current measurements [7].

### *3.2 Proposed optimal estimator*

In order to estimate the parameters for a TL, all methods previously discussed require 3-phase voltage and current PMU phasors measurements from both terminals of the line. We will refer the phasor measurements taken at one time instant as one sample. The relationships among the values in one sample are described by (40). That is, 12 real equations are needed to describe the interdependencies of one sample. Using multiple samples of PMU measurements, the underdetermined set of equations, (40), becomes an overdetermined set of equations and state estimation techniques can be used to estimate the TL's parameters.

Assuming N PMU samples taken at N distinct time instants are available, we can set up 12*N equations, causing $Z, H, \theta$ to have the following dimensions respectively: 12*N by 1, 12*N by 18, 18 by 1. And finally we arrive at a typical over-determined linear state estimation problem. Using the unbiased least square estimator [8], the best estimation of the unknown vector $\theta$ is found to be:

$$\theta = (H^T \cdot H)^{-1} \cdot H^T \cdot Z \quad (41)$$

Once $\theta$ is known, the phase impedance parameters can be retrieved easily based on (38). After the phase impedance matrix is calculated, applying the sequence transformation matrix will yield the sequence impedance parameters. The equations used here are shown below:

$$Z_S = A^{-1} Z_{abc} A = A^{-1} (y_P^{-1}) A \tag{42}$$

$$B_S = A^{-1} B_{abc} A \tag{43}$$

For noisless PMU measurements, only two PMU samples are needed to calculate the impedance parameters accurately. Otherwise, multiple PMU samples are necessary to increase the redundancy so that linear estimation theory and bad data detection and elimination can be conducted. Classical methods for bad data detection can be found in [8-10].

In the classical methods, bad data are identified primarily based on the study of the model scaled residuals. First, all the data are utilized to conduct the parameters' estimation. The scaled residuals for each sample of measurements are checked. The samples containing large unexpected errors will have scaled residuals much larger than the average variance, which actually can help to identify those bad PMU data samples. After removing these bad data, the linear estimation technique is utilized once again to calculate the impedance.

Case studies showing how this method compares with the single and double measurement methods will be presented in the next major section. In the next subsection we show how this method can be modified to account for mutual coupled TL's.

### *3.3 Application of the model to mutually coupled TL's*

Many transmission lines are located adjacent to each other, or are under-built with lower voltage transmission/distribution lines. These adjacent or under-build lines induced voltages which skew the PMU measurements. These induced voltages cause the estimated impedance parameters to be erroneous when the methods utilizing only the positive components are used. And these errors are considerable if the mutual coupling is significant and the TL is not fully transposed [7]. We show next that the method proposed here may be modified to compensate for the voltages induce by the mutually coupled TL's.

The effect of mutual coupling is to induce a voltage in series with the transmission line conduction voltage drop. To compensate for this effect, this induced voltage must be subtracted from the voltage measurements. This induced voltage can be calculated based on the mutual inductance and the current phasor measurements of the transmission line inducing the voltage on the line to be studied. Once the induced voltages for all three phases are estimated, these induced voltages can be subtracted from the voltage drop across the transmission line by making a small change to the left side of equation (28) as:

$$y_P \cdot (\overline{U}_{abc}^{S} - \overline{U}_{abc}^{R} - \overline{U}_{abc}^{induced}) = \overline{I}_{abc}^{S} - j \cdot \frac{1}{2} \cdot B_{abc} \cdot \overline{U}_{abc}^{S} \tag{44}$$

where $U_{abc}^{induced}$ is the estimated induced voltage phasors calculated from PMU measurements from other transmission lines.

This idea was tested and satisfactory results were obtained in [7]. Due to the space limitation, detailed information and test results will not be presented in this paper.

## 4. CASE STUDIES

In this section, 3 case studies are presented to demonstrate the procedures and validate the effectiveness of the proposed method. The Alternative Transients Program (ATP) is employed to build a 3-phase transmission line model and to generate synchrophasor measurements [11]. One 230 kV transmission line is simulated using typical physical parameters (e.g., tower geometry, conductor type) obtained from Salt River Project (SRP) [APPENDIX B]. The transmission lines are built in ATP using the LCC objects. The LCC subroutine automatically calculates the impedance parameters of the TL's once the physical parameters are input. These impedance parameters calculated by the LCC subroutine are taken as the true parameters of the TL's, and are used as reference values to be compared with the calculated parameters.

Because the method we propose requires redundant measurements obtained from different loading conditions, a time varying load is modeled on the receiving-end of the transmission lines. The load curve is set to be sinusoidal with a period of 24 hours. It varies between one 20% to 80% of the maximum line capacity. To make sure that each sample represents a different load condition, samples are taken every five minutes.

Another objective of this section is to demonstrate the limitation of the positive sequence TL model when applied to parameter estimation of untransposed TL's. In order to do that, two algorithms proposed in [3] and [4] are employed to calculate the impedance parameters for an untransposed TL with unbalanced loading conditions. The results obtained from these two algorithms are compared with the results from the proposed method.

### *4.1 Case for the fully transposed TL*

To validate the proposed parameter estimation method, the following experiment was conducted. A 14.5-km-long *completely transposed* TL with the parameters shown in Appendix A was modeled in ATP with a diurnal sinusoidally varying load as described above. Voltage and current phasors for 200 different loading conditions were sampled from the ATP output and no noise was added to these measurements. These measurements served as input to the method proposed and the parameters that characterize the transmission line are estimated. Since the line is fully transposed, there are six non-zero parameters that represent the TL; these are: the positive, zero and negative sequence series impedances and shunt susceptances. Applying the proposed method to the simulated synchrophasor measurements, the results shown in Table I are obtained.

As expected, Table I shows that accurate estimates are achieved by the proposed method. The other mutual impedances (e. g., $Z_{10}$ , $Z_{12}$ ) not shown are calculated by the algorithm but are essentially zero as expected.

### *4.2 Case for the untransposed TL*

In section 2, this paper illustrates the limitation of the positive sequence model and the methods based on this model. In this subsection, the methods proposed in reference [3] and [4] are tested with synchrophasor data obtained from one untransposed TL with unbalanced load. This experiment was the same as that described in Section 4.1 except the line was untransposed and the load was unbalanced. In the simulation with 14% unbalanced loads were applied to the TL. The degree of unbalance is defined in this paper as the ratio of negative sequence current to the positive sequence current ($\frac{I_2}{I_1} \times 100\ \%$). The results for these two methods are shown in Table II and Table III, which show inaccurate impedance parameters are calculated for untransposed line with unbalanced load when the single and double measurement methods are used.

Table I. Optimal estimator for fully transposed line with proposed method

| ***Quantity*** | **Reference values** | **Optimal estimates** | **\|Error(%) in R\|** | **\|Error(%) in X or B\|** |
|---|---|---|---|---|
| $Z_1(\Omega)$ | 0.8839 +j 6.9188 | 0.8837+j6.9192 | 0.022% | 0.006% |
| $Z_2(\Omega)$ | 0.8839 + j6.9188 | 0.8837 +j6.9192 | 0.022% | 0.006% |
| $Z_0(\Omega)$ | 3.3455 +j22.8057 | 3.3530 +j22.8157 | 0.224% | 0.044% |
| $B_1(S)$ | j5.0198e-05 | j5.0195e-05 | - | 0.006% |
| $B_2(S)$ | j5.0198e-05i | j5.0195e-05 | - | 0.006% |
| $B_0(S)$ | j2.7633e-05i | j2.7625e-05 | - | 0.029% |

Table II. Calculated parameters for untransposed line with unbalanced load using the double measurement method in [3]

| ***Quantity*** | **Reference values** | **Calculated values** | **\|Error(%)\|** |
|---|---|---|---|
| $R_1 (\Omega)$ | 0.8839 | 1.2983 | 46.88% |
| $X_1 (\Omega)$ | 6.9188 | 7.7830 | 12.49% |
| $B_1 (\Omega)$ | 5.0188e-05 | 4.5394e-05 | 9.55% |

Table III. Calculated parameters for untransposed line with unbalanced load using single measurement method in [4]

| ***Quantity*** | **Reference values** | **Calculated values** | **\|Error(%)\|** |
|---|---|---|---|
| $R_1 (\Omega)$ | 0.8839 | 1.0240 | 15.85% |
| $X_1 (\Omega)$ | 6.9188 | 7.7151 | 11.51% |
| $B_1 (\Omega)$ | 5.0188e-05 | 4.5402e-05 | 9.54% |

Table IV. Calculated parameters for untransposed line with unbalanced load using proposed method

| ***Quantity*** | **Reference values** | **Optimal estimates** | **\|Error(%) in R\|** | **\|Error(%) in X or B\|** |
|---|---|---|---|---|
| $Z_0 (\Omega)$ | 3.3455+j22.8057 | 3.3455+j22.8070 | 0% | 0.006% |
| $Z_1 (\Omega)^*$ | 0.8839+j6.9188 | 0.8839+j6.9194 | 0% | 0.009% |
| $Z_{01} (\Omega)$ | 0.2213-j0.1396 | 0.2213-j0.1396 | 0% | 0% |
| $Z_{02} (\Omega)$ | -0.2054-j0.1305 | -0.2055-j0.1306 | 0.049% | 0.077% |
| $Z_{10} (\Omega)$ | -0.2054-j0.1305 | -0.2055-j0.1306 | 0.049% | 0.077% |
| $Z_{12} (\Omega)$ | -0.4369+j0.2524 | -0.4370+j0.2524 | 0.023% | 0% |
| $Z_{20} (\Omega)$ | 0.2213-j0.1396 | 0.2213-0.1396 | 0% | 0% |
| $Z_{21} (\Omega)$ | 0.4371+j0.2522 | 0.4371+j0.2522 | 0% | 0% |
| $B_0 (S)$ | 2.7630e-05 | 2.7625e-05 | - | 0.018% |
| $B_1 (S)$ | 5.0188e-05 | 5.0192e-05 | - | 0.008% |
| $B_{01} (S)^*$ | 1.5805e-06 | 1.5807e-06 | - | 0.013% |
| $B_{02} (S)$ | 1.5805e-06 | 1.5805e-06 | - | 0% |
| ${}^*B_{12} (S)$ | -1.6869e-06 | -1.6875e-06 | - | 0.036% |

$^*Z_1 = Z_2, B_1 = B_2, B_{01} = B_{10}, B_{02} = B_{20}, B_{12} = B_{21}$

The performance of the proposed method is evaluated for the same TL with the same degree of 14% unbalance. Since the TL is untransposed, the impedance matrix calculated is a full matrix. The results are shown in Table IV, where close agreement can be found between the reference values and the optimal estimates. It should be noted that, with the proposed method, not only the diagonal elements but also the non-diagonal elements in the impedance matrix can be calculated accurately.

### *4.3 Case for untransposed TL with noise in the measurements*

PMU data are time tagged with accuracy of better than 1 microsecond and magnitude accuracy that is better than 1%. However, this potential performance is not achieved in actual field installations due to unbiased random measurement noise and biased errors from instrumentation channels [2], [12]. Figure 2 shows the magnitude and phase angle of voltage phasor measurements taken from a PMU. As shown in the two plots, there are spikes and complex measurement noise behavior in the phasor measurements. The spikes in the plots represent bad data, which may be removed through bad data detection techniques [8-10]. However, many times, it is very difficult and may even be impossible to filter out all data corrupted with noise. As will be shown, the single and double measurement methods are sensitive to this noise, where as the optimal parameter estimation algorithm is able to reduce the effects of this noise on the parameter estimation.

Another complication brought on by noise is that its effect on parameter estimation is more pronounced on shorter transmission lines. To examine the effect of noise on parameter estimation in general and its effect on line length in particular, we conducted experiments to explore the relationship between the accuracy of the estimated parameters and the length of transmission line under noisy measurement conditions. The noise considered was Gauss noise with 0 mean and 1% standard deviation. For a certain length of transmission line, we ran the ATP simulation of the untransposed line and obtained N samples of the synchrophasor measurements. These N samples of measurements are referred to as one set of measurements. Then we added different noise vectors to the synchrophasor measurements to form M different sets of measurements. And then we applied the proposed method and the single and double measurement methods to each set of the noisy measurements to estimate the impedances of the TL. By applying these methods to the M different sets of noisy measurements, we generated different estimates of the impedance parameters for the transmission line described in APPENDIX B. Assuming that the impedance parameters estimated this way were normally distributed, we calculated the 95% confidence interval of these parameters, that is, the range in which our estimated parameter fell 95% of the time. We repeated this process for different lengths of transmission lines so that we could obtain the confidence intervals of the impedance parameters for different lengths of TL's.

The standard deviation of the error in the R parameter versus length of the transmission line is shown in Figure 3. Similar plots for X and B are shown in Figure 4 and Figure 5.

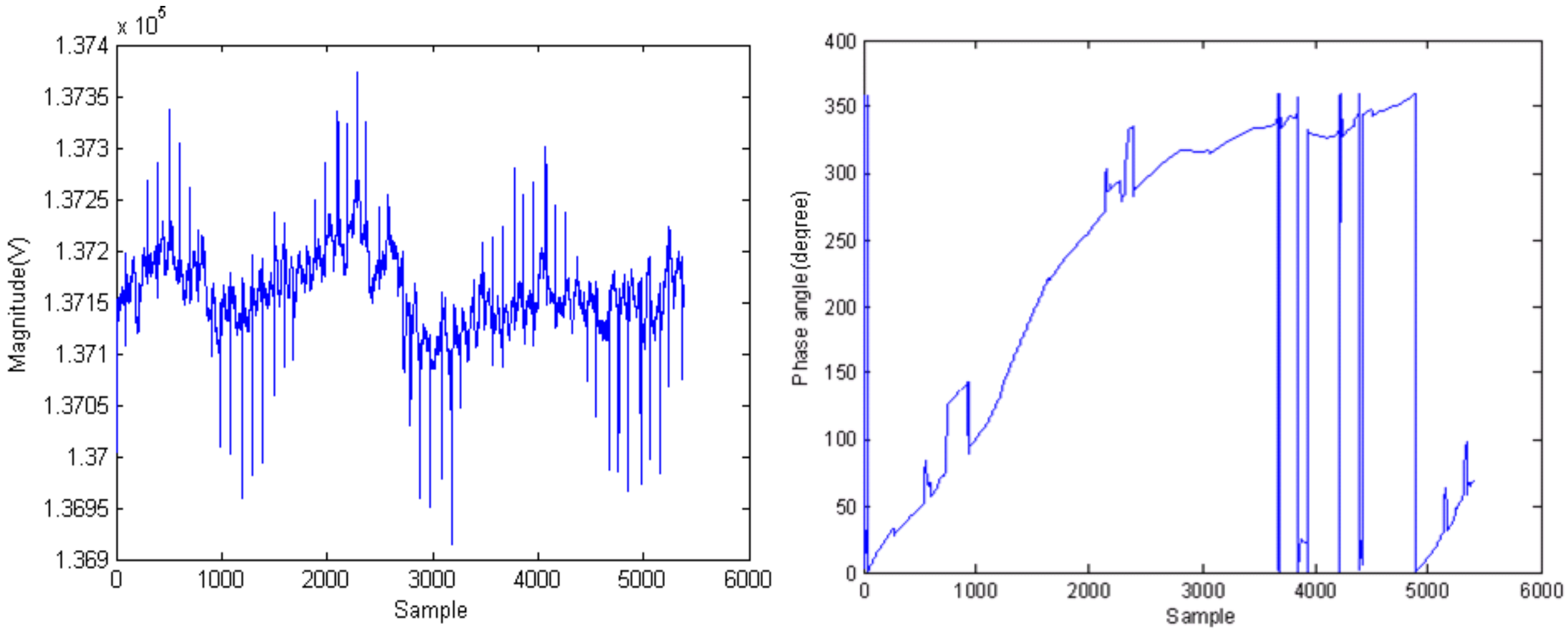

Figure 2 Recorded magnitude and phase angle of a voltage phasor from a PMU

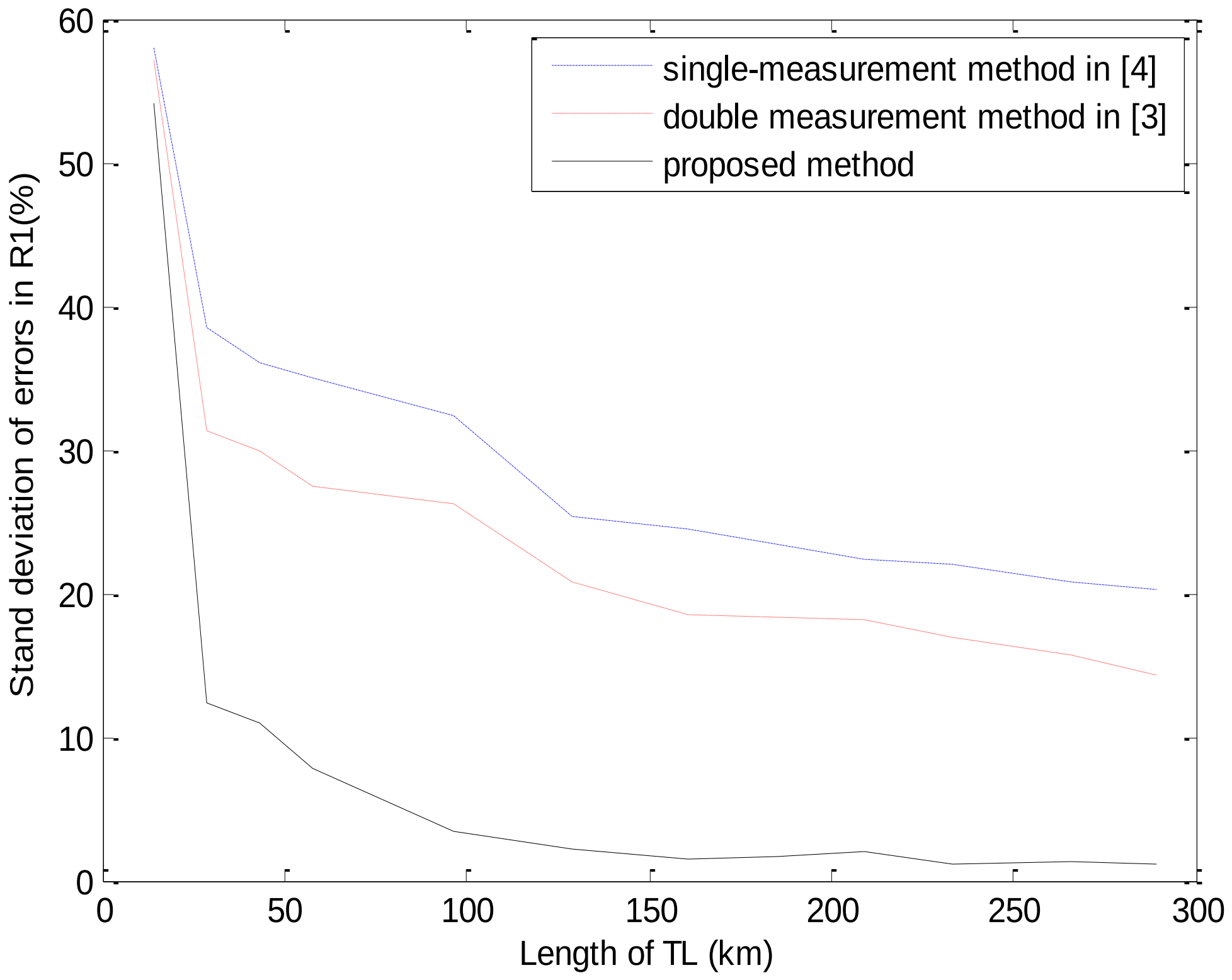

Figure 3. Standard deviation of Errors in calculated $R_1$ for TL's of different lengths

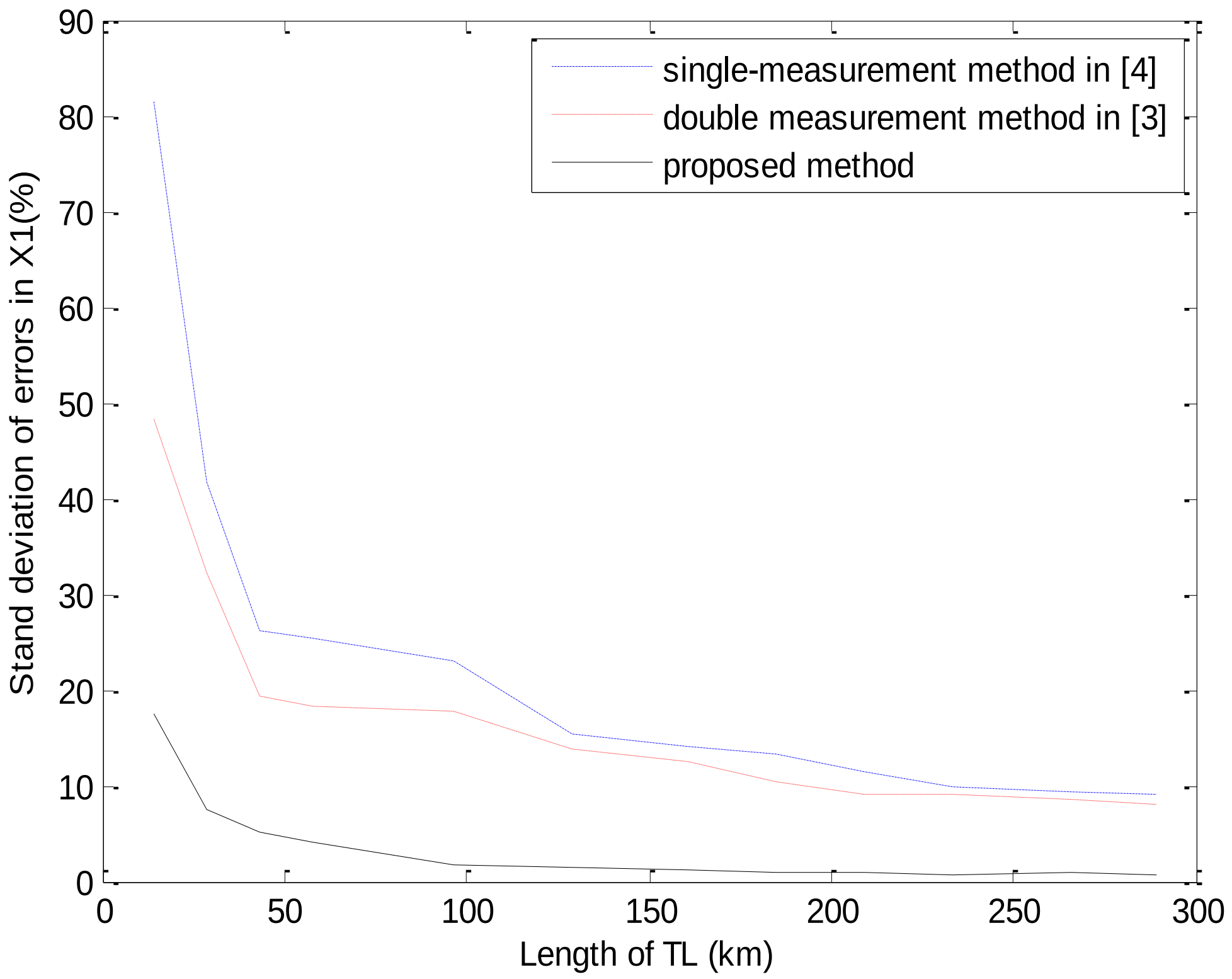

Figure 4. Standard deviation of errors in $X_1$ for TL's of different lengths

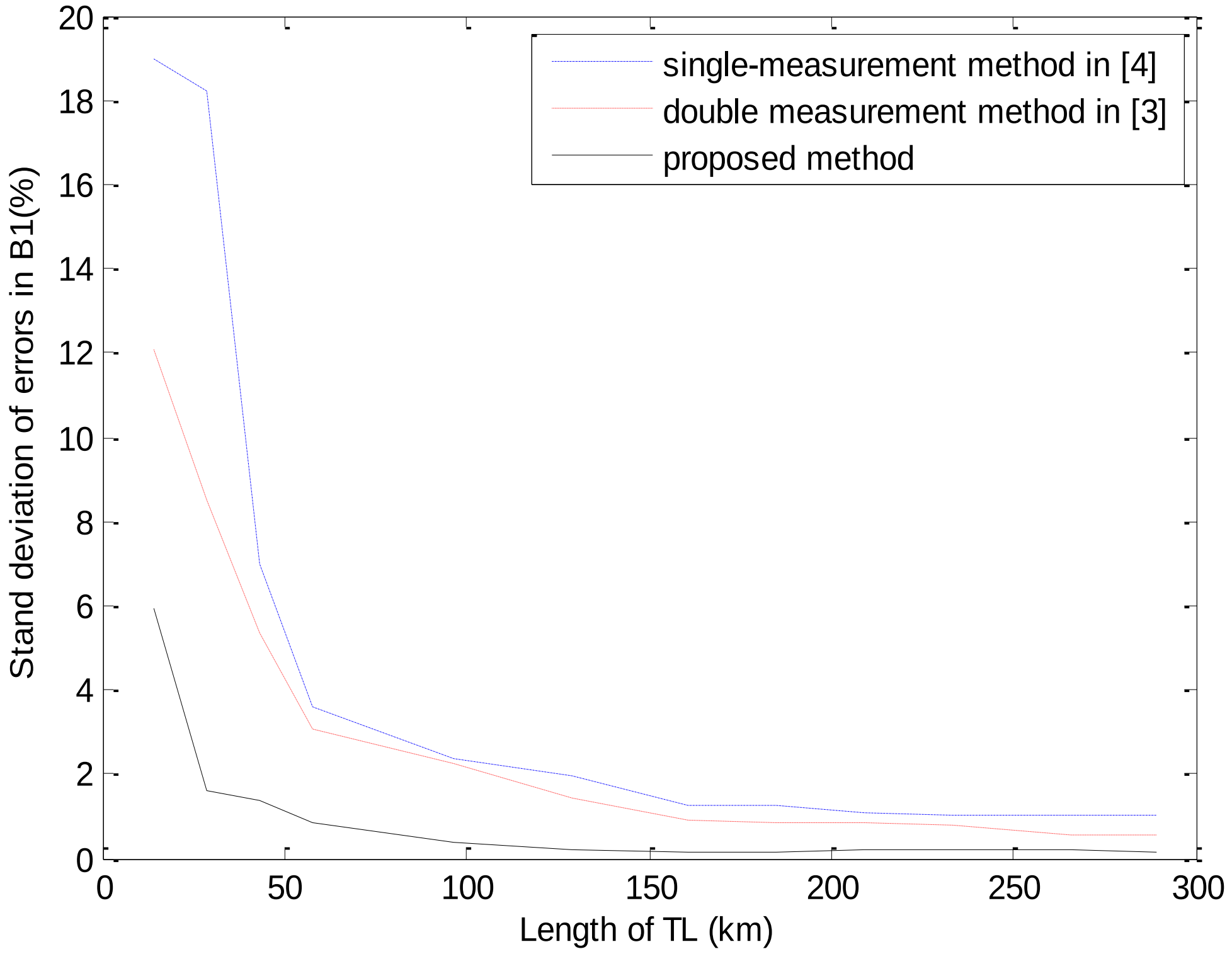

Figure 5. Standard deviation of errors in $B_1$ for TL's of different lengths

Figure 3-Figure 5 show the standard deviation (SD) of errors in calculated impedance parameters as a function of TL lengths when the PMU measurements are corrupted by Gauss noise (0, 1%). Figure 4 shows that for a 14.5-km long transmission line (an in-service line for which we were asked by Salt River Project to estimate the parameters), with Gauss noise (0, 1%) added, the SD's of errors in calculated $X_1$ reach about 80%, 50%, and 18% of the true value by the single measurement method, the double measurement method and the proposed method, respectively. As the length of the transmission line increases, the SD's of the errors decrease and the SD's calculated by the proposed method are always much smaller than the single or double measurement methods. If the transmission line is longer than 150 kilometers, with the same sets of noise added, the deviation in the error of the calculated series reactance will be smaller than 1% for the proposed method while the SD's by the other two methods remain at about 14%. These figures clearly show the advantage of the proposed method over the other two methods. And the proposed method is particularly superior to the other two methods for short transmission lines.

## 5. CONCLUSION

This objective of this paper is to compare various methods for estimating the parameters for TL's (under transposed and untransposed configurations and under balanced and unbalanced conditions) using synchrophasor measurements. Particularly of interest, is the application of these methods to short TL's. In this paper, it is shown that the positive sequence model of a TL has limitations when used to construct the defining equations for TL parameter estimation. Parameter estimation methods based on the positive sequence model alone are shown to perform particularly poorly when applied to short untransposed and unbalanced transmission lines. This paper proposed a novel optimal parameter estimation method to accurately estimate the impedance parameters for a general transposed/untransposed TL with balanced or unbalanced loading. This method is based on the linear estimation theory and assumes availability of redundant PMU measurements. As stated earlier, bad data detection method can also be utilized to identify and remove bad PMU data samples. Several case studies are conducted and the performance of this optimal estimator is compared with other methods. The encouraging results shown here are for a typical transmission line configuration.

# 6. LIST OF ABBREVIATIONS AND SYMBOLS

## *6.1 Abbreviations*

| | |
|---|---|
| ATP | Alternative Transients Program |
| GPS | Global Positioning System |
| LCC | line/cable constant |
| PMU | Phasor Measurement Unit |
| SD | standard deviation |
| SRP | Salt River Project |
| TL | transmission line |

## *6.2 Symbols*

| | |
|---|---|
| $A$ | sequence transformation matrix |
| $B_{abc}$ | 3-phase shunt susceptance matrix (3 by 3) |
| $B_S$ | 3-sequence shunt susceptance matrix (3 by 3) |
| $B_x\,(x = a, ab ...)$ | self shunt susceptance of one phase or mutual shunt susceptance between phases; element of $B_{abc}$ |
| $B_x\,(x = 0,1,01 ...)$ | shunt susceptance of one sequence or mutual shunt susceptance between different sequences; elements of $B_S$ |
| $\det(Z_{abc})$ | determinant of $Z_{abc}$ |
| $\overline{I}_{abc}^{S(R)}$ | sending (receiving) end phase current vector (3 by 1) |
| $\overline{I}_{x}^{S(R)}\,(x = a, b, c)$ | sending (receiving) end phase current |
| $\overline{I}_{012}^{S(R)}$ | sending (receiving) end sequence current vector (3 by 1) |
| $\overline{I}_{x}^{S(R)}\,(x = 0,1,2)$ | sending (receiving) end sequence current |
| $\mathrm{Im}(\cdot)$ | imaginary part of a number |
| $\mathrm{Re}(\cdot)$ | real part of a number |
| $\overline{U}_{abc}^{S(R)}$ | sending (receiving) end phase voltage vector (3 by 1) |
| $U_{x}^{S(R)}\,(x = a, b, c)$ | sending (receiving) end phase voltage |
| $\overline{U}_{012}^{S(R)}$ | sending (receiving) end sequence voltage vector (3 by 1) |
| $U_{x}^{S(R)}\,(x = 0,1,2)$ | sending (receiving) end sequence voltage |
| $y_P$ | inverse of $Z_{abc}$ (3 by 3) |
| $y_x\,(x = a, b, ab, ...)$ | elements in $y_P$ |
| $Y_{abc}$ | 3-phase shunt admittance matrix (3 by 3) |
| $Y_x\,(x = a, ab, ...)$ | self shunt admittance of one phase or mutual shunt susceptance between phases; element of $Y_{abc}$ |
| $Y_{012}$ | 3-sequence shunt admittance matrix (3 by 3) |
| $Y_x\,(x = 0,1,01,...)$ | self shunt susceptance of one sequence or mutual shunt susceptance between different sequences; elements of $Y_{012}$ |
| $\overline{U}_{abc}^{induced}$ | vector (3 by 1) comprised of induced voltages (due to mutual coupling) on a 3-phase transmission line |
| $Z_{abc}$ | 3-phase series impedance matrix (3 by 3) |
| $Z_x\,(x = a, ab, ...)$ | self series impedance of one phase or mutual series impedance between phases; element of $Z_{abc}$ |
| $Z_{012}$ | 3-sequence series impedance matrix (3 by 3) |
| $Z_x\,(x = 0,1,01,...)$ | self series impedance of one sequence or mutual series impedance between different sequences; elements of $Z_{abc}$ |
| $Z_{abc}^{-1}$ | inverse of $Z_{abc}$ |

## 7. APPENDIX A

Complex equations (31)~(36) are expanded into 12 real equations. By defining $y_x = G_x + j \cdot T_x\,(x = a, b, c, ab, bc, or \quad ac)$, these 12 equations can be written as:

$$\begin{aligned} &G_a \operatorname{Re}(\Delta U_a) - T_a \operatorname{Im}(\Delta U_a) + G_{ab} \operatorname{Re}(\Delta U_b) - T_{ab} \operatorname{Im}(\Delta U_b) + G_{ac} \operatorname{Re}(\Delta U_c) - T_{ac} \operatorname{Im}(\Delta U_c) \\ &- \operatorname{Re}(I_a^S) - \frac{1}{2} B_a \operatorname{Im}(U_a^S) - \frac{1}{2} B_{ab} \operatorname{Im}(U_b^S) - \frac{1}{2} B_{ac} \operatorname{Im}(U_c^S) = 0 \end{aligned} \quad \text{(A. 1)}$$

$$\begin{aligned} &G_{ab} \operatorname{Re}(\Delta U_a) - T_{ab} \operatorname{Im}(\Delta U_a) + G_b \operatorname{Re}(\Delta U_b) - T_b \operatorname{Im}(\Delta U_b) + G_{bc} \operatorname{Re}(\Delta U_c) - T_{bc} \operatorname{Im}(\Delta U_c) \\ &- \operatorname{Re}(I_b^S) - \frac{1}{2} B_{ab} \operatorname{Im}(U_a^S) - \frac{1}{2} B_b \operatorname{Im}(U_b^S) - \frac{1}{2} B_{bc} \operatorname{Im}(U_c^S) = 0 \end{aligned} \quad \text{(A. 2)}$$

$$\begin{aligned} &G_{ac} \operatorname{Re}(\Delta U_a) - T_{ac} \operatorname{Im}(\Delta U_a) + G_{bc} \operatorname{Re}(\Delta U_b) - T_{bc} \operatorname{Im}(\Delta U_b) + G_c \operatorname{Re}(\Delta U_c) - T_c \operatorname{Im}(\Delta U_c) \\ &- \operatorname{Re}(I_c^S) - \frac{1}{2} B_{ac} \operatorname{Im}(U_a^S) - \frac{1}{2} B_{bc} \operatorname{Im}(U_b^S) - \frac{1}{2} B_c \operatorname{Im}(U_c^S) = 0 \end{aligned} \quad \text{(A. 3)}$$

$$\frac{1}{2} B_a \operatorname{Im}(U_a^S + U_a^R) + \frac{1}{2} B_{ab} \operatorname{Im}(U_b^S + U_b^R) + \frac{1}{2} B_{ac} \operatorname{Im}(U_c^S + U_c^R) + \operatorname{Re}(\Delta I_a) = 0 \quad \text{(A. 4)}$$

$$\frac{1}{2} B_{ab} \operatorname{Im}(U_a^S + U_a^R) + \frac{1}{2} B_b \operatorname{Im}(U_b^S + U_b^R) + \frac{1}{2} B_{bc} \operatorname{Im}(U_c^S + U_c^R) + \operatorname{Re}(\Delta I_b) = 0 \quad \text{(A. 5)}$$

$$\frac{1}{2} B_{ac} \operatorname{Im}(U_a^S + U_a^R) + \frac{1}{2} B_{bc} \operatorname{Im}(U_b^S + U_b^R) + \frac{1}{2} B_c \operatorname{Im}(U_c^S + U_c^R) + \operatorname{Re}(\Delta I_c) = 0 \quad \text{(A. 6)}$$

$$\begin{aligned} &G_a \operatorname{Im}(\Delta U_a) + T_a \operatorname{Re}(\Delta U_a) + G_{ab} \operatorname{Im}(\Delta U_b) + T_{ab} \operatorname{Re}(\Delta U_b) + G_{ac} \operatorname{Im}(\Delta U_c) + T_{ac} \operatorname{Re}(\Delta U_c) \\ &- \operatorname{Im}(I_a^S) + \frac{1}{2} B_a \operatorname{Re}(U_a^S) + \frac{1}{2} B_{ab} \operatorname{Re}(U_b^S) + \frac{1}{2} B_{ac} \operatorname{Re}(U_c^S) = 0 \end{aligned} \quad \text{(A. 7)}$$

$$\begin{aligned} &G_{ab} \operatorname{Im}(\Delta U_a) + T_{ab} \operatorname{Re}(\Delta U_a) + G_b \operatorname{Im}(\Delta U_b) + T_b \operatorname{Re}(\Delta U_b) + G_{bc} \operatorname{Im}(\Delta U_c) + T_{bc} \operatorname{Re}(\Delta U_c) \\ &- \operatorname{Im}(I_b^S) + \frac{1}{2} B_{ab} \operatorname{Re}(U_a^S) + \frac{1}{2} B_b \operatorname{Re}(U_b^S) + \frac{1}{2} B_{bc} \operatorname{Re}(U_c^S) = 0 \end{aligned} \quad \text{(A. 8)}$$

$$G_{ac}\ \mathrm{Im}(\Delta U_a) + T_{ac}\ \mathrm{Re}(\Delta U_a) + G_{bc}\ \mathrm{Im}(\Delta U_b) + T_{bc}\ \mathrm{Re}(\Delta U_b) + G_c\ \mathrm{Im}(\Delta U_c) + T_c\ \mathrm{Re}(\Delta U_c)$$
$$- \mathrm{Im}(I_c^S) + \frac{1}{2}B_{ac}\ \mathrm{Re}(U_a^S) + \frac{1}{2}B_{bc}\ \mathrm{Re}(U_b^S) + \frac{1}{2}B_c\ \mathrm{Re}(U_c^S) = 0 \tag{A. 9}$$

$$\frac{1}{2}B_a\ \mathrm{Re}(V_a^S + V_a^R) + \frac{1}{2}B_{ab}\ \mathrm{Re}(V_b^S + V_b^R) + \frac{1}{2}B_{ac}\ \mathrm{Re}(V_c^S + V_c^R) - \mathrm{Im}(\Delta I_a) = 0 \tag{A. 10}$$

$$\frac{1}{2}B_{ab}\ \mathrm{Re}(U_a^S + U_a^R) + \frac{1}{2}B_b\ \mathrm{Re}(U_b^S + U_b^R) + \frac{1}{2}B_{bc}\ \mathrm{Re}(U_c^S + V_c^R) - \mathrm{Im}(\Delta I_b) = 0 \tag{A. 11}$$

$$\frac{1}{2}B_{ac}\ \mathrm{Im}(U_a^S + U_a^R) + \frac{1}{2}B_{bc}\ \mathrm{Im}(U_b^S + U_b^R) + \frac{1}{2}B_c\ \mathrm{Re}(U_c^S + U_c^R) - \mathrm{Im}(\Delta I_a) = 0 \tag{A. 12}$$

where Re(.) and Im(.) yield the real and imaginary parts of the input argument, respectively.

## 8. APPENDIX B

The physical parameters (e.g., tower geometry, conductor type) of the transmission line used in the simulations as described in section 4 are shown in Table B.I. The soil resistivity is assumed to be 100 $ohm \cdot m$ in the simulation.

Table. B. I. Transmission line physical parameters used in simulations

| Phase. No | Reactance (ohm/mile AC) | Outer radius of the conductor (inch) | Conductor resistance at Freq (with no skin effect) (ohm/mile AC) | Horizontal distance from the center of the tower (feet) | Vertical bundle height at tower (feet) | Vertical bundle height at mid-span (feet) |
|---|---|---|---|---|---|---|
| 1 | 0.39 | 0.598 | 0.0982 | 0 | 48 | 38 |
| 2 | 0.39 | 0.598 | 0.0982 | 0 | 66 | 56 |
| 3 | 0.39 | 0.598 | 0.0982 | 0 | 84 | 74 |

*system frequency is 60 Hz; all phase conductors are single-bundled conductors

ACKNOWLEDGEMENTS

The authors thank Salt River Project for the support in carrying out this research work.